\documentclass[12pt]{article}
\usepackage{amssymb}
\usepackage{amsmath,,calrsfs}
\usepackage[dvips]{epsfig}

\newcommand{\be}{\begin{equation}}
	\newcommand{\ee}{\end{equation}}
\def\bea{\begin{eqnarray}}
	\def\eea{\end{eqnarray}}

\newcommand{\bn}{\begin{eqnarray}}
	\newcommand{\en}{\end{eqnarray}}

\newcommand{\p}{\partial}

\newcommand{\nn}{\nonumber}

\newcommand{\no}{\noindent}

\newcommand{\s}{\,\,\,\,}
\def\bea{\begin{eqnarray}}
	\def\eea{\end{eqnarray}}

\newcommand{\beq}{\begin{eqnarray}}
	\newcommand{\eeq}{\end{eqnarray}}
\begin{document}
	
	\title{\textbf{Master Actions and Helicity Decomposition for Spin-4 Models in $3D$}}
	\author{Elias L. Mendon\c ca \footnote{elias.leite@unesp.br}, R. Schimidt Bittencourt \footnote{raphael.schimidt@unesp.br}\\
		\textit{{UNESP - Campus de Guaratinguet\'a - DFI} }\\
		\textit{{Av. Dr. Ariberto Pereira da Cunha, 333} }\\
		\textit{{CEP 12516-410 - Guaratinguet\'a - SP - Brazil.} }\\
	}
	\date{\today}
	\maketitle
	
	\begin{abstract}
		The present work introduces a master action which interpolates between four self-dual models, $SD(i)$, describing massive spin-4 particles in $D=2+1$ dimensions. These models are designated by $i=1,2,3$ and $4$, representing the order in derivatives. Our results show that the four descriptions are quantum equivalents by comparing their correlation functions, up to contact terms. This is an original result since that a proof of quantum equivalence among these models have not been demonstrated in the literature. Besides, a geometrical approach is demonstrated to be a useful tool in order to describe the third and fourth order in derivatives models. On the other hand, the construction of the master action relies on the introduction of mixing terms, which must be free of particle content. Here, we demonstrate how the helicity decomposition method can be used in order to verify the absence of particle content of such terms, ensuring the proper usability of the master action technique. This kind of result can be very useful in situations where we do not have access to the higher spin-projection basis which would allow the analysis to be handle by explicitly calculating the propagator and subsequently analyzing its poles. 
	\end{abstract}
	\newpage
	
	\section{Introduction}

	An explicit Lagrangian for a free massive field of arbitrary spin was first introduced by Singh and Hagen \cite{SH1, SH2} in terms of totally symmetric, single traceless rank-$s$, $s-2$, $s-3$, ... $0$ tensors. One can then observe that in this description, $s-2$ lower rank fields are introduced in order to eliminate the spurious degrees of freedom on-shell, and because of this, they are called auxiliary fields. Such an elimination of spurious DOF is achieved through the so-called Fierz-Pauli conditions. Fronsdal and Fang \cite{Fronsdal, Fang} studied the massless limit of the Singh-Hagen theory and concluded that, except for the rank-$s-2$ field, all the other auxiliaries decouple from the Lagrangian. The remaining two fields, the rank $s$ and $s-2$ fields, can then be combined into a single symmetric, double traceless field. The resulting Lagrangian becomes gauge invariant, with the gauge parameter being a rank-$s-1$ symmetric and traceless tensor. When working with the gauge invariant description, the restrictions on the gauge parameter can be used as a guide for constructing geometrical objects analogous to the Riemann, Ricci, and Einstein tensors, as it has being done, for example, in the spin-3 case \cite{deserdamour}. See also a simple generalization for the spin-4 case \cite{nges4} and other interesting geometrical approaches by \cite{XBek1, XBek2, Xbek3, Henneaux, Dario, Hohm} where the reader can find a fully systematized and generalized analysis of the work made by T. Damour and S. Deser  on the spin-3 field \cite{deserdamour}. Some of the massive models for higher-spin particles can be elegantly written in terms of such objects.
	
	A particular feature of 3D space-times is the existence of topologically massive gauge theories, which can be achieved for both Yang-Mills and gravitational descriptions by adding gauge-invariant topological mass terms to the Lagrangian. This procedure was introduced in the seminal works by Deser, Jackiw, and Templeton \cite{djt1, djt2}. In the same vein, several papers have suggested generalizations of these models for different spins, both bosonic and fermionic, known as self-dual (SD) descriptions, as seen in the bosonic cases in the works of P.K.Townsend, K. Pilch, P. Nieuwenhuizen, C. Aragone and A. Khoudeir \cite{Pilch, aragone1, aragone2}.
	
	Regarding the particle content of these models, recently the authors of \cite{Zino} have confirmed that the self-dual descriptions suggested by Aragone and Khoudeir actually propagate only one massive particle with the desired spin without any ghosts in the spectrum. Indeed, in the original work {\cite{aragone2}} the authors have already demonstrated through the study of the equations of motion that the massive, spin-4, first order self-dual model is free of ghost and correctly describe the massive spin-4 field. In this work, we start precisely from this model in order to demonstrate that it is quantum equivalent to higher-derivative, massive, and gauge-invariant descriptions that we have previously obtained in {\cite{nges4}}. Additionally, we show that a geometric description can elegantly be used in order to describe the third and fourth-order in derivatives models. Such demonstration of quantum equivalence is an important and original result, because the master action technique is, as far as we know, the unique dualization procedure which is able to guarantee effectively the quantum equivalence of dual descriptions.

   More specifically, in this work, starting with the first-order self-dual model introduced in \cite{aragone2} in order to describe a massive spin-4 mode in $D=2+1$ we construct a master action that can interpolate with other three self-dual descriptions of second, third, and fourth orders in derivatives. The first-order self-dual model is free of the propagation of spurious degrees of freedom, thanks to the addition of auxiliary fields. We demonstrate that, during the master action interpolation process, such auxiliary fields are recombined step by step. The construction of the master action requires a crucial ingredient, the mixing terms, which must be trivial i.e. free of particle content. In order to verify the triviality of such mixing terms we demonstrate that one can use the helicity decomposition technique \cite{deserh, andringa, townsendh}.
   
   The paper is structured as follows: Section 2 provides an overview of the basic notation, including the spin-4 field and a useful shorthand notation without indices that is used when working with the master action and certain ``geometrical'' objects. Section 3 is dedicated to introducing the master action and is divided into three sub-sections, where the reader is guided through the verification of the spectrum of the first order self-dual model (our ``starting point'' model), the introduction of the so called trivial mixing terms and finally the construction of the master action for spin-4 self-dual models. In Section 4, we have demonstrated that the master action we suggest can be used in order to demonstrate that higher derivative self-dual descriptions are quantum equivalent to the first order self-dual model. This is an interesting result, since in the literature such descriptions are only classically related. It is also shown that it is possible and advantageous to switch from the notation using partially symmetric fields to the notation using totally symmetric fields  where geometrical structures naturally emerges from the dualization procedure.

	\section{Notation} 
	Throughout this work, we use the mostly plus flat metric $(-,+,+)$. The spin-4 self-dual descriptions in $D=2+1$ can be written in the frame-like or in the totally symmetric approach. In the frame-like approach \cite{Vasiliev}, the spin-4 field is described by $\omega_{\mu(\alpha\beta\gamma)}$, which is symmetric and traceless in its Lorentz-like indices between parentheses. However, the trace can be defined as $\eta^{\mu\alpha}\omega_{\mu(\alpha\beta\gamma)}=\omega_{\beta\gamma}$. However notice that $\eta^{\beta\gamma}\omega_{\beta\gamma}=0$. It is worth mentioning that, for readers familiar with the notation in terms of differential forms, the field $\omega_{\mu(\alpha\beta\gamma)}$ actually consists of a 1-form valued in the Lorentz algebra representation for symmetric and traceless rank-3 tensors. In this context, the index $\mu$ corresponds to a world index while the other indices are Lorentz indices. However, we want to emphasize that, in this particular setting, we are working in flat space and then  we do not distinguish between indices inside and outside the parentheses.
	
	The 21 independent components of $\omega_{\mu(\alpha\beta\gamma)}$ can be represented by the following feasible decomposition:
	\be \omega _{\mu(\beta\gamma\lambda)}= \phi_{\mu\beta\gamma\lambda}+\frac{1}{3}\eta_{\mu(\beta}\phi_{\gamma\lambda)}-\frac{1}{3}\eta_{(\beta\gamma}\phi_{\lambda)\mu}+ \,\, \epsilon_{\mu\rho(\beta} \tilde{\chi}^{\rho}_{\,\,\,\gamma\lambda)}.\label{deco}\ee
	Where the symmetrisation convention we have adopted along the text is unnormalized. In the right-hand side of (\ref{deco}), we have the 14 components of the totally symmetric double traceless field $\phi_{\mu\beta\gamma\lambda}$ plus the $7$ components of the totally symmetric traceless field $\tilde{\chi}_{\mu\beta\gamma}$, which correctly balance the number of independent components on both sides of (\ref{deco}). In this work, the numerical coefficients used in the decomposition (\ref{deco}) are chosen in such a way that the terms in the frame-like notation coincide exactly with those in the totally symmetric approach suggested by \cite{deserdamour}. Finally, the last term in (\ref{deco}) is indeed a symmetry of the second, third, and fourth order in derivative terms (see these terms below). Please notice that this is not an irreducible decomposition; instead, it represents a possible decomposition in terms of fully symmetric fields. We employ it for the sole purpose of transitioning from the notation involving dreibeins $\omega$ to the fully symmetric notation expressed in terms of $\phi$."
	
	The master action notation we use through this work is exactly based on the same idea we have used in \cite{dualdesc}, then, we define a shorthand notation similar to that used there to express the zero, first, second and third - order in derivative terms, respectively given by:
	\bea \int \,\, (\omega^2)&\equiv& \int d^3x\,\, (\omega_{\mu\beta}\omega^{\mu\beta}-\omega_{\mu(\beta\gamma\lambda)}\omega^{\beta(\mu\gamma\lambda)}),\\
	\int \,\, \omega\cdot d\omega&\equiv&\int d^3x\,\,\epsilon^{\mu\nu\alpha}\omega_{\mu(\beta\gamma\lambda)}\p_{\nu}\omega_{\alpha}^{\s(\beta\gamma\lambda)},\label{csl}\\
	\int\,\, \omega\cdot d\Omega(\omega)&\equiv&\int d^3x\,\,\epsilon^{\mu\nu\alpha}\omega_{\mu(\beta\gamma\lambda)}\p_{\nu}\Omega_{\alpha}^{\s(\beta\gamma\lambda)}\left\lbrack \xi(\omega)\right\rbrack,\label{ehl}\\
	\int \,\, \Omega(w)\cdot d\Omega(w)&\equiv&\int d^3x\,\,\epsilon^{\mu\nu\alpha}\Omega_{\mu(\beta\gamma\lambda)}\left\lbrack \xi(\omega)\right\rbrack\p_{\nu}\Omega_{\alpha}^{\s(\beta\gamma\lambda)}\left\lbrack \xi(\omega)\right\rbrack\label{tcs}\eea
	
	\no where the symbol $.d \equiv \epsilon^{\mu\nu\alpha}\p_{\nu}$, while $\Omega_{\rho(\alpha\beta\gamma)}$ \footnote{It may be possible that there is an identification between the symbol $\Omega$ and the generalized Christoffel symbol proposed by \cite{Wit}}. has the same symmetry properties of the field $\omega_{\rho(\alpha\beta\gamma)}$ and is given by:
	\bea  \Omega_{\rho(\alpha\beta\gamma)}\left\lbrack \xi(\omega)\right\rbrack&\equiv&\xi_{\rho(\alpha\beta \gamma)}-\frac{1}{2}\Big(\xi_{\alpha(\rho\beta \gamma)}+\xi_{\beta(\rho\alpha\gamma)}+\xi_{\gamma(\rho\beta\alpha)}\Big)\nn\\
	&-&\frac{1}{8}\Big(\eta_{\rho\alpha}\xi_{\beta \gamma}+\eta_{\rho\beta}\xi_{\alpha\gamma}+\eta_{\rho\gamma}\xi_{\beta\alpha}\Big)\cr\cr
	&+&\frac{1}{4}\Big(\eta_{\beta\alpha}\xi_{\rho\gamma}+\eta_{\gamma\beta}\xi_{\alpha\rho}+\eta_{\alpha\gamma}
	\xi_{\beta\rho}\Big).\label{OM1}\eea
	
	\no Here, $\xi_{\mu(\beta\gamma\lambda)}\equiv E_{\mu\nu}\omega^{\nu}_{\s(\beta\gamma\lambda)}$, where the operator $E_{\mu\nu}\equiv\epsilon_{\mu\nu\alpha}\p^{\alpha}$ is the explicit expression for the derivative symbol, $.d$. In the notation of the master action, where indices are omitted, we will use simply $\Omega(\omega)$ instead of $\Omega[\xi(\omega)]$ for clearer calculations. However, when a field or combination of fields is present in the same tensorial structure offered by the symbol $\Omega$, but without the derivatives contained in $\xi$, we denote the $\Omega$-symbol with a tilde, e.g. $\tilde{\Omega}(a)$ is the expression (\ref{OM1}) with $a$ replacing $\xi$ i.e. free of derivatives. 
	
	Throughout the manipulations with the master action, we will frequently use the self-adjoint property of the symbol $\Omega$, which one can easily verify from its definition and it is given by:
	
	\be \int \, A\cdot d \Omega (B) = \int \, B\cdot d \Omega (A),\ee
	\be
	\int d^3x\, A_{\mu(\alpha\beta\gamma)} \Omega^{\mu(\alpha\beta\gamma)}(B) = \int d^3x\, B_{\mu(\alpha\beta\gamma)} \Omega^{\mu(\alpha\beta\gamma)}(A)
	\ee
	\no for any $A$ and $B$. While the notation we have introduced above may appear similar to that used in the context of differential forms, it is important to note that they are, in fact, distinct. We will revisit this similarity in the conclusion, as it plays a crucial role in our attempt to extend our findings to fields of arbitrary rank-s.
	
	The frame-like and the totally symmetric descriptions can be related each other by mean of the decomposition (\ref{deco}). Substituting (\ref{deco}) in (\ref{ehl}) we have:
	\bea \int \, \omega\cdot d\Omega(\omega)
	&=& -\frac{1}{2} \int d^3x\,\,\, \phi_{\mu\nu\lambda\beta}\,\mathbb{G}^{\mu\nu\lambda\beta}(\phi),\label{secondsym}\eea
	\no where we have introduced the ``Einstein tensor'' $\mathbb{G}$ given by:
	\be \mathbb{G}_{\mu\nu\lambda\beta}\equiv \mathbb{R}_{\mu\nu\lambda\beta}-\frac{1}{2}\eta_{(\mu\nu}\mathbb{R}_{\lambda\beta)}.\label{G}\ee
	
	\no whith:
	\be \mathbb{R}_{\mu\nu\lambda\beta}= \Box \phi_{\mu\nu\lambda\beta}-\p_{(\mu}\p^{\alpha}\phi_{\alpha\nu\lambda\beta)}+\p_{(\mu}\p_{\nu}\phi_{\lambda\beta)},\label{Ricci} \ee
	
	\no and:
	\be \mathbb{R}_{\lambda\beta}=2\left[\Box \phi_{\lambda\beta}-\p^{\mu}\p^{\alpha}\phi_{\mu\alpha\lambda\beta}+\frac{1}{2}\p_{(\beta}\p^{\alpha}\phi_{\alpha\lambda)}\right].\label {R}\ee

	\no It is important to emphasize that the objects defined in equations (\ref{G}), (\ref{Ricci}), and (\ref{R}) are simply spin-4 analogues of the geometric objects defined in General Relativity. As explained in \cite{deserdamour}, higher spin gauge theories differ from the spin-2 case due to the lack of a unified geometric relationship between the background and the dynamical field. This is why the term "geometry" is presented in quotation marks throughout this text. Substituting equation (\ref{deco}) into the third-order term given by equation (\ref{tcs}), we have:
		
	\bea \int\,\, \Omega(w)\cdot d\Omega(w)
	&=& \frac{1}{16} \int d^3x\,\,\mathbb{C}_{\mu\nu\gamma\lambda}(\phi)\mathbb{G}^{\mu\nu\gamma\lambda}(\phi).\label{thirdsym}\eea

\no The operator $\mathbb{C}$, a symmetrized curl, is given by:
\be \mathbb{C}_{\mu\nu\gamma\lambda}(\phi)\equiv - E_{(\mu}^{\,\,\,\,\,\,\beta}\phi_{\beta\nu\gamma\lambda)}.\label{C}\ee

\no Both symbols $\mathbb{G}$ and $\mathbb{C}$ were constructed based on the assumptions in \cite{deserdamour} and possess the following algebraic properties: The operator $\mathbb{G}^{\mu\nu\lambda\beta}(\phi)$ is self-adjoint, meaning that $\int \psi \mathbb{G}(\phi) = \int \phi \mathbb{G}(\psi)$ for any symmetric fields $\psi$ and $\phi$. This property is crucial for deriving the equations of motion and for the interpolation process in the master action technique. Similarly, the operator $\mathbb{C}_{\mu\nu\gamma\lambda}(\phi)$ is also self-adjoint. Additionally, it can be verified that the operators $\mathbb{C}$ and $\mathbb{G}$ commute with each other, in the sense that $\int \phi \mathbb{C}[\mathbb{G}(\phi)] = \int \phi \mathbb{G}[\mathbb{C}(\phi)]$.

	\section{Constructing the Master Action}
	
	As we have demonstrated in some of our previous works, see for example \cite{dualdesc} and subsequent works, the key ingredients in order to construct master actions are:  1st) To have a {\bf first order in derivatives self-dual model free of ghosts}. 2nd) To have at hands the so-called {\bf trivial mixing terms} which will be plugged to the self-dual model. Together, these ingredients allow us to construct the so called master action. In the subsection 3.1 we revisit the equations of motion for the first order self-dual model confirming that it really describes the massive spin-4 propagation. In the subsection 3.2, using the helicity decomposition method we check the triviality of the suggested mixing terms. Finally in the subsection 3.3 we construct the master action and demonstrate that it has the same particle content as the first order self-dual model.
	\subsection{ The first order self-dual model as a starting point}
	We have seen what are the basic ingredients in order to construct a master action to be used in the interpolation process among the different self-dual descriptions. Let us use as a starting point the first-order self-dual model suggested by \cite{aragone2}, which, making use of our notation is given by:
	\bea S_{SD}^{(1)}&=& \int \left[\frac{m}{2}\omega\cdot d\omega +\frac{m^2}{2}(\omega^2)+ m^2\,\,\omega_{\mu\nu}U^{\mu\nu}\right]+
	S^1_{aux}[U,H,V].\label{SD(1)}\eea
	Such model, was originally suggested  as a higher spin generalization of the self-dual descriptions of spins 1 and 2, in $D=2+1$, given by \cite{Pilch,aragone1} aiming an investigation of the higher-spin structure of topologically massive theories. As we can observe the model can not avoid the presence of auxiliary fields, namely $U^{\mu\nu}$, $H^{\mu}$ and $V^{\mu}$ and the explicit expression of the auxiliary action $S_{aux}^1$ is given by:
	\begin{eqnarray}
		\label{acao21}
		S^1_{aux}&=&\int d^3x  \ \ \bigg[-\frac{3m}{4}\epsilon^{\rho\mu\nu}U_{\rho\alpha}\partial_\mu U_{\nu}^{\alpha} -\frac{3m^2}{2}(U^2 - U_{\mu\nu}U^{\nu\mu})-\frac{8m}{9}\epsilon^{\mu\nu\beta}H_\mu\partial_\nu H_\beta\cr\cr
		&-& \frac{9m}{20}\epsilon^{\mu\nu\beta}V_\mu\partial_\nu V_\beta + \frac{32m^2}{9}H_\mu H^\mu - \frac{9m^2}{5}V_\mu V^\mu + m^2H^\mu V_\mu-\frac{9m}{5}U\partial_\mu V^\mu + \frac{22m^2}{5} U^2\bigg].\nn\\
	\end{eqnarray}
	All the numerical coefficients in the auxiliary action have been determined in \cite{aragone2} from an analysis of the equations of motion. Let us understand how the model is connected with the auxiliary fields as well as to recover some aspects of the equations of motion in a cleared way. 
	
	The original rank-4 field $\omega_{\mu(\beta\gamma\lambda)}$ may be algebraically irreducible decomposed as:
	\be \omega_{\mu(\beta\gamma\lambda)}= {\bar \omega}_{(\mu\beta\gamma\lambda)}+\frac{5}{21}\eta_{\mu(\beta}{ \omega}_{\gamma\lambda)}-\frac{2}{21}{ \omega}_{\mu(\beta}\eta_{\gamma\lambda)}+\epsilon_{\mu(\beta}^{\s\s\,\, \alpha}\bar{H}_{(\alpha\gamma\lambda))},\label{irred}\ee
	
	\no where the barred tensors in the right hand side stands for totally symmetric, single traceless fields and $\omega_{\mu\nu}$ is the trace of $\omega_{\mu(\beta\gamma\lambda)}$ which by its turn is also traceless i.e.: $\eta^{\mu\nu}\omega_{\mu\nu}=0$ as we have mentioned in the notation section. Remember that in order to describe the spin-4 mode we need a rank-4 totally symmetric, traceless and transverse field surviving in the equations of motion of the given model. Besides, one can easily check that the number of components on both sides of (\ref{irred}) is correctly balanced.  
	
	Now, we notice that in (\ref{SD(1)}) the trace of $\omega_{\mu(\beta\gamma\lambda)}$, i.e. $\omega_{\gamma\lambda}$ is coupled to the auxiliary rank-2 nonsymmetric field $U^{\gamma\lambda}$, and such field may be algebraically irreducible decomposed as:
	\be U^{\gamma\lambda}=\bar{U}^{(\gamma\lambda)}+\epsilon^{\gamma\lambda\alpha}H_{\alpha}+\frac{1}{3}\eta^{\gamma\lambda}U\label{decou}\ee
	
	So, we have learned that in the auxiliary action (\ref{acao21}) the auxiliary fields $U$ and $H^{\mu}$ actually corresponds to the irreducible parts of the nonsymmetric $U^{\gamma\lambda}$ field. Additionally, an auxiliary vector $V^{\mu}$ is also introduced.
	
	Now, let us take the equations of motion for the action (\ref{SD(1)}) in order to verify that all the auxiliary fields vanishes on shell:
	\bea K^{\mu(\alpha\beta\gamma)}\equiv\frac{\delta S_{SD(1)}^{(1)}}{\delta \omega_{\mu(\alpha\beta\gamma)}}&=&-\sqrt{\Box}\hat{E}^{\mu\nu}\omega_{\nu}^{\,\,\,(\alpha\beta\gamma)}+\frac{m}{3}(\eta^{\mu\alpha}\omega^{\beta\gamma}+\eta^{\mu\beta}\omega^{\alpha\gamma}+\eta^{\mu\gamma}\omega^{\beta\alpha})\nn\\
	&-&\frac{m}{3}(\omega^{\alpha(\mu\beta\gamma)}+\omega^{\beta(\mu\alpha\gamma)}+\omega^{\gamma(\mu\beta\alpha)})+\frac{m}{3}(\eta^{\mu\alpha}U^{\beta\gamma}+\eta^{\mu\beta}U^{\alpha\gamma}+\eta^{\mu\gamma}U^{\beta\alpha})\nn\\
	&-&\frac{2m}{15}(U^{\mu\alpha}\eta^{\beta\gamma}+U^{\mu\beta}\eta^{\alpha\gamma}+U^{\mu\gamma}\eta^{\beta\alpha})
	\eea
	where we have used $\hat{E}^{\mu\nu}=E^{\mu\nu}/\sqrt{\Box}$. For the field $U_{\mu\alpha}$ we have:
	\bea F^{\mu\alpha} \equiv\frac{\delta S_{SD(1)}^{(1)}}{\delta U_{\mu\alpha}}&=&m\omega^{\mu\alpha}+\frac{3}{2}\sqrt{\Box}\hat{E}^{\mu\nu}U_{\nu}^{\,\,\,\alpha}+3m(\eta^{\mu\alpha}U-U^{\mu\alpha})-\frac{8}{9}(\p^{\mu}H^{\alpha}-\p^{\alpha}H^{\mu})\nn\\
	&-&\frac{32}{9}m\epsilon^{\mu\alpha\nu}H_{\nu}-\frac{1}{2}m\epsilon^{\mu\alpha\nu}V_{\nu}-\frac{9}{5}\eta^{\mu\alpha}\p_{\nu}V^{\nu}+\frac{44m}{5}\eta^{\mu\alpha}U\label{equ}\eea
	Notice that, in (\ref{decou}) one can invert $H^{\mu}$ in terms of $U^{\mu\nu}$ as $H^{\mu}=-\epsilon^{\mu\alpha\beta}U_{\alpha\beta}/2$. We have used this information in order to obtain (\ref{equ}). Finally, let us take the equation of motion for $V^{\mu}$.
	\bea G^{\mu}\equiv \frac{\delta S_{SD(1)}^{(1)}}{\delta V_{\mu}}=\frac{9}{10}\sqrt{\Box}\hat{E}^{\mu\nu}V_{\nu}-\frac{18}{5}mV^{\mu}+mH^{\mu}+\frac{9}{5}\p^{\mu}U. \eea
	
	While the spin-4 field is described by the totally symmetric, traceless and transverse part of $\omega_{\mu(\alpha\beta\gamma)}$, one can demonstrate that the action (\ref{SD(1)}) does not propagate (surprisingly) any mode of spin-3, which could be described by the transverse part of $\p_{\mu}\bar{\omega}^{(\mu\alpha\beta\gamma)}$ and $\bar{H}^{(\alpha\beta\gamma)}$. This can be understood if we take the following manipulations of the equations of motion: $\p_{\alpha}K^{\alpha(\mu\beta\gamma)}=0$ and $\epsilon_{(\mu}^{\s\rho\lambda}K_{\rho(\lambda\beta\gamma))}=0$ which results respectively in:
	\be \p_{\mu}\omega_{\beta\gamma}+\p_{\beta}\omega_{\mu\gamma} +\p_{\gamma}\omega_{\beta\mu} - \p^{\alpha}\omega_{\mu(\alpha\beta\gamma)} - \p^{\alpha}\omega_{\beta(\alpha\mu\gamma)}- \p^{\alpha}\omega_{\gamma(\alpha\beta\mu)}= \frac{2}{5}\eta_{\mu(\beta}\p^{\alpha}\bar{U}_{(\alpha\gamma))}-\p_{(\mu}\bar{U}_{(\beta\gamma))}\\\label{primeira}\ee
	\be 4m\bar{H}_{(\mu\beta\gamma)}=\frac{2}{5}\eta_{\mu(\beta}\p^{\alpha}\bar{U}_{(\alpha\gamma))}-\p_{(\mu}\bar{U}_{(\beta\gamma))}, \label{segunda}\ee
	\no where we have seen that the spin-3 variables are in fact proportional to curls of the spin-2 variable. Then their pure spin-3 parts have necessarily to vanish and one can conclude that the action (\ref{SD(1)}) is not propagating spin-3 modes. We have been analyzing first, the equations of motion of the spin-3 sector because they are constrained, i.e. the right hand side of (\ref{primeira}) and (\ref{segunda}) are the same. These will facilitate the verification of the lower spin sectors in the subsequent lines.
	
	Now, let us show that the spin zero modes, described by the seven scalars $\p_{\mu}\p_{\nu}\p_{\beta}\p_{\gamma} \bar{\omega}^{(\mu\nu\beta\gamma)}$, $\p_{\mu}\p_{\nu}\omega^{\mu\nu}$, $\p_{\mu}\p_{\nu}\p_{\beta}\bar{H}^{(\mu\beta\gamma)}$, $\p_{\mu}\p_{\nu}\bar{U}^{(\mu\nu)}$, $U$, $\p_{\mu}H^{\mu}$ and finally $\p_{\mu}V^{\mu}$ get vanished on shell. The seven equations which describes these seven variables are: 
	\bea \p_{\mu}\p_{\nu}\p_{\beta}\p_{\gamma} K^{\mu(\nu\beta\gamma)}&=&0 \\
	  \p_{\mu}\p_{\nu}\p_{\beta}(\tilde{K}^{\mu(\nu\beta)}+\tilde{K}^{\nu(\mu\beta)}+\tilde{K}^{\beta(\nu\mu)})&=&0\\
	  \eta_{\mu\nu}\p_{\alpha}\p_{\beta}K^{\mu(\nu\alpha\beta)}&=&0\\
	  \p_{\mu}\tilde{F}^{\mu}&=&0\\
	  \eta_{\mu\nu}F^{\mu\nu}&=&0\\
	  \p_{\mu}\p_{\nu}F^{\mu\nu}&=&0\\
	  \p_{\mu}G^{\mu}&=&0\eea
	\no  where we have defined $\tilde{K}^{\mu(\nu\beta)} \equiv \epsilon^{\mu}_{\s\alpha\gamma}K^{\alpha(\gamma\nu\beta)}$ and $\tilde{F}^{\mu}\equiv \epsilon^{\mu}_{\s\,\alpha\beta}F^{\alpha\beta}$. These seven equations may be written in matrix form as:
	\bea\begin{pmatrix}
		0 & 0 & 4m & 9\Box/5 & 0 & 0 & 0\\
		-5 & 20\Box/70 & 0 & 3\Box & 0 & 0 & 0\\
		0 & 4m/3 & -4 & 7m/5 & 0 & 0 & 0\\
		0 & 0 & 0 & 3/2 & -\Box & 10m/9 & m\\
		0 & 0 & 0 & 0 & 54m/5 & 1 & -9/5\\
		0 & m & 0 & -3m & 54m\Box/5 & 0 & -9\Box/5\\
		0 & 0 & 0 & 0 & 9\Box/5 & m & -18m/5\\
	\end{pmatrix} 
 \begin{pmatrix}
	\p_{\mu}\p_{\nu}\p_{\beta}\p_{\gamma} \bar{\omega}^{(\mu\nu\beta\gamma)} \\
	\p_{\mu}\p_{\nu}\omega^{\mu\nu} \\
	\p_{\mu}\p_{\nu}\p_{\beta}\bar{H}^{(\mu\beta\gamma)} \\
	\p_{\mu}\p_{\nu}\bar{U}^{(\mu\nu)} \\
	U \\
	\p_{\mu}H^{\mu} \\
	\p_{\mu}V^{\mu}\\
\end{pmatrix}=0\eea
	\no Now, in order to guarantee that the model is free of the propagation of any spin-zero mode we need to verify that the determinant of the $7 \times 7$ matrix above is different than zero and free of d'Alembertians, and in fact one can easily check that, by using any calculator software.
	
	We then proceed to the verification of the spin-1 sector which is described in terms of the transverse part of the six variables $\p_{\nu}\p_{\beta}\p_{\gamma} \bar{\omega}^{(\mu\nu\beta\gamma)}$, $\p_{\nu}\omega^{\mu\nu}$, $\p_{\nu}\p_{\beta}\bar{H}^{(\mu\beta\gamma)}$, $\p_{\mu}\p_{\nu}\bar{U}^{(\mu\nu)}$, $H^{\mu}$ and finally $V^{\mu}$. In order to verify that they are non-propagating we take the following six vector equations already projected in terms of the transverse variables:
	
	\bea P_{\pm \s \beta}^{(1)\alpha}[\p_{\mu}\p_{\nu}\p_{\gamma} K^{\mu(\nu\gamma\beta)}]&=&0 \\ 
	P_{\pm \s \beta}^{(1)\alpha}[\p_{\mu}\p_{\nu}(\tilde{K}^{\mu(\nu\beta)}+\tilde{K}^{\nu(\mu\beta)}+\tilde{K}^{\beta(\nu\mu)})]&=&0\\
	P_{\pm \s \beta}^{(1)\alpha}[\eta_{\mu\nu}\p_{\alpha}K^{\mu(\nu\alpha\beta)}]&=&0\\
	P_{\pm \s \beta}^{(1)\alpha}[\p_{\mu}F^{\mu\beta}]&=&0\\
	P_{\pm \s \beta}^{(1)\alpha}[\tilde{F}^{\beta}]&=&0\\
	P_{\pm \s \beta}^{(1)\alpha}[G^{\beta}]&=&0\eea
	
	\no where the spin $\pm 1$ projection operator, $P_{\pm \s \beta}^{(1)\alpha}$, is given by:
	\be P_{\pm \s \beta}^{(1)\alpha}=\frac{1}{2}\left(\theta^{\alpha}_{\beta}\pm \hat{E}^{\alpha}_{\beta}\right) \ee
	\no whit $\theta_{\beta}^{\alpha}=\left(\delta_{\beta}^{\alpha}-\p_{\beta}\p^{\alpha}/\Box\right)$. In matrix form the six equations of motion are written as:
	{\scriptsize \bea\begin{pmatrix}
	
		 3 & -50\Box/21 & 4m\pm 5\sqrt{\Box} & 0 & 0 & 0\\
		 0 & 0 & 4m & 8\Box/5 & 0 & 0\\
		 0 & (4m\pm\sqrt{\Box})/3  & -4 & 7m/5 & 0 & 0\\
		 0 & m & 0 & -3m & (-8\Box \mp5m\sqrt{\Box})/9 & \mp m\sqrt{\Box}/2\\
		 0 & 0 & 0 & 3/2 & (20m\pm 5\sqrt{\Box})/18& m\\
		 0 & 0 & 0 & 0 & m & (-36m\pm9\sqrt{\Box})/10\\
	\end{pmatrix} 
	\begin{pmatrix}
		P_{\pm \s \beta}^{(1)\alpha}[\p_{\mu}\p_{\nu}\p_{\gamma} \bar{\omega}^{(\mu\nu\gamma\beta)}] \\
		P_{\pm \s \beta}^{(1)\alpha}[\p_{\mu}\omega^{\mu\beta}] \\
		P_{\pm \s \beta}^{(1)\alpha}[\p_{\mu}\p_{\nu}\bar{H}^{(\mu\nu\beta)}] \\
		P_{\pm \s \beta}^{(1)\alpha}[\p_{\mu}\bar{U}^{(\mu\beta)}] \\
		P_{\pm \s \beta}^{(1)\alpha}[H^{\beta}] \\
		P_{\pm \s \beta}^{(1)\alpha}[V^{\beta}]\\
	\end{pmatrix}=0\eea}
	
	\no Then, we have demonstrated that the spin-1 variables are vanishing on-shell, since the determinant of the $6\times 6$ matrix is different than zero and free of d'Alembertians. 
	
	Now, let us focus on the spin-2 variables, namely, those described by the transverse parts of: $\p_{\mu}\p_{\nu}\bar{\omega}^{(\mu\nu\beta\gamma)}$, $\p_{\mu}\bar{H}^{(\mu\beta\gamma)}$, $\omega^{\beta\gamma}$ and $\bar{U}^{(\beta\gamma)}$ whose the dynamics is governed by the following equations $\p_{\mu}\p_{\nu}K^{\mu(\nu\beta\gamma)}=0$, $F^{(\mu\alpha)}=0$, $\eta_{\mu\nu}K^{\mu(\nu\beta\gamma)}=0$ and $\p^{\mu}\epsilon_{(\mu}^{\s\rho\lambda}K_{\rho(\lambda\beta\gamma))}=0$. However, notice that the first equation is exactly the same as the last equation, thanks to what we have observed in the equations (\ref{primeira}) and (\ref{segunda}) which have the same right hand side. Then we just need to care with the following three equations given by:
	
	\be 4m\p^{\mu}\bar{H}_{(\mu\beta\gamma)}=-\Box\bar{U}_{(\beta\gamma)}\label{1}\ee
	\be \omega^{\mu\nu}=3\bar{U}^{\mu\nu}-\frac{3\sqrt{\Box}}{4m}\hat{E}^{(\mu}_{\s\,\,\alpha}\bar{U}^{\alpha\nu)}\label{2}\ee
	\be -\frac{\sqrt{\Box}}{3}(\hat{E}^{\mu}_{\s\,\,\alpha}\omega_{\beta\mu}+\hat{E}^{\mu}_{\s\,\,\beta}\omega_{\alpha\mu})-4\p^{\mu}\bar{H}_{(\mu\alpha\beta)}+\frac{4}{3}m\omega_{\alpha\beta}+\frac{7}{5}m\bar{U}_{(\alpha\beta)}=0,\label{3}\ee

	\no where we have used the irreducible decompositions (\ref{irred}) and (\ref{decou}). Substituting (\ref{1}) and (\ref{2}) in (\ref{3}) one can verify that $\bar{U}^{(\mu\nu)}=0$ which automatically imply that $\omega^{\mu\nu}=0$, $\bar{H}_{(\alpha\mu\nu)}=0$ and then, consequently, $\p^{\mu}\bar{\omega}_{(\mu\beta\gamma\lambda)}=0$.   
	
	The immediate consequence of the last results is that the spin-4 sector become described only by the following equation of motion:
	\be E^{\mu}_{\s\nu}\bar{\omega}^{(\nu\alpha\beta\gamma)}-m\bar{\omega}^{(\mu\alpha\beta\gamma)}=0 \label{Pauli}\ee
	\no where one can easily check that through the application of the operator $E_{\beta\mu}$ to the same it results in the Klein-Gordon equation, confirming than that the model (\ref{SD(1)}) really describes a unique massive spin-4 mode without any lower spin propagations. In a work in progress, about the Lorentz generators for arbitrary spins, we are demonstrating that the equation (\ref{Pauli}) may be written as a Pauli-Lubanski equation. In the next subsection we study the particle content of the so called mixing terms, which then will be added to the action (\ref{SD(1)}) studied in the present subsection.

	\subsection{The triviality of the mixing terms}
	The other key ingredient in order to constructing the master actions are the mixing terms. During calculations, after some of the shifts, they can be completely decoupled and once integrated, one can get rid of them from the Lagrangians. Then, since the mixing terms get decoupled after the manipulations and are free of particle content one can conclude by the quantum equivalence of the interpolated models through the master action procedure. Here, we use three mixing terms, given by: (\ref{csl}), (\ref{ehl}) and (\ref{tcs}). The first term, (\ref{csl}) is clearly without particle content, because it has the same structure of the spin-1 Chern-Simons term and one can check that the solution of its equation of motion is indeed pure gauge. The second term, given by (\ref{ehl}), is actually the massless Fronsdal theory which is known to be trivial in $D=2+1$, however, the demonstration of triviality to the third order in derivatives term, given by (\ref{tcs}), is not that easy. Here, we show that the decomposition in helicities method, used for example in \cite{deserh, andringa, townsendh} can be used in a relatively simple way, in order to verify the particle content of models and isolated terms. First we use the second order massless Fronsdal term as a warm up exercise and then we study the most complicated case of the third order in derivatives term. Another possible way of analysis is through the determination of the propagator, but, it would be necessary to have in hands the complete basis of spin-projectors as we have done for example in \cite{elmrs1, elmrs2} in the spin-3 case.
	\subsubsection{The second order mixing term}

	The second order Eistein-Hilbert like term (massless Fronsdal), is given by:
	
	\be {\cal L}^{(2)}= \frac{1}{2}\phi_{\mu\nu\lambda\beta}\,\mathbb{G}^{\mu\nu\lambda\beta}(\phi)\label{secondorderterm}\ee
	\no where the superscript indicates the order in derivatives, is written in terms of the totally symmetric double traceless field $\phi_{\alpha\nu\lambda\beta}$, which in $D=2+1$ dimensions, have $14$ independent components. The action is invariant under the reparametrizations:
	\be \delta \phi_{\alpha\nu\lambda\beta}=\p_{(\alpha}\tilde{\xi}_{\nu\lambda\beta)},\label{second}\ee
	\no where the gauge parameter $\tilde{\xi}_{\nu\lambda\beta}$ is completely symmetric and traceless, thus, in $D=2+1$ it has $7$ independent components. This allow us to fix $14-7=7$ constraint equations, which can be written as:
	\be \p_i\phi_{ijk\mu}=0.\label{constraint}\ee
	Where Latin indices stands for spatial indices. Such constraint equations can be fixed in the action level, since they are {\it complete}, in  the sense of that explored by \cite{moto1, moto2}. To demonstrate that they are complete we must be able to invert the gauge parameter in terms of the field without any ambiguities. After some tedious calculations one can indeed verify such inversion:
	\be \xi_{jk\mu}=-\frac{\p_i\phi_{ijk\mu}}{\nabla^2}+\frac{\p_m\p_n\p_{(j}\phi_{mnk\mu)}}{2\nabla^4}-\frac{\p_m\p_n\p_p\p_{(j}\p_k\phi_{mnp\mu)}}{3\nabla^6}+\frac{\p_m\p_n\p_p\p_q\p_{j}\p_k\p_{\mu}\phi_{mnpq}}{4\nabla^8}.\ee
	
	\no Where $\nabla^2=\p_i\p_i$. Noticing that the double traceless condition allow us the elimination of $\phi_{0000}=2\phi_{00ii}-\phi_{iijj}$, one can use the following helicity decomposition:
	\be \phi_{0000}=2\varphi_3-\nabla^4\theta\quad; \quad\phi_{000i}=\hat{\p}_i\gamma+\p_i\Gamma\label{decomp1}\ee
	\be \phi_{00ij}=(\hat{\p}_i\p_j+\p_i\hat{\p}_j)\varphi_1+\left(\p_i\p_j-\frac{\delta_{ij}\nabla^2}{2}\right)\varphi_2+\frac{\delta_{ij}\varphi_3}{2}\ee
	\be \phi_{0ijk}=\hat{\p}_i\hat{\p}_j\hat{\p}_k\psi\quad;\quad \phi_{ijkl}=\hat{\p}_i\hat{\p}_j\hat{\p}_k\hat{\p}_l\theta\ee
	where $\hat{\p}_i\equiv \epsilon_{ij}\p_j$, while the traces are given by:
	\be \phi_{00}=-\varphi_3+\nabla^4\theta\quad;\quad \phi_{0i}=-\hat{\p}_i\gamma-\p_i\Gamma+\hat{\p}_i\nabla^2\psi\ee
	\be \phi_{ij}=-(\hat{\p}_i\p_j+\p_i\hat{\p}_j)\varphi_1-\left(\p_i\p_j-\frac{\delta_{ij}\nabla^2}{2}\right)\varphi_2-\frac{\delta_{ij}\varphi_3}{2}+\hat{\p}_i\hat{\p}_j\nabla^2\theta. \label{decomp2}\ee
	
	\no One can easily verify that the number of independent components of the tensors in the left hand side of each decomposition equation coincides with the number of scalar fields $(\varphi_1, \varphi_2, \varphi_3, \theta, \gamma, \Gamma,\psi)$ we have in the right hand side of them. Opening the Lagrangian (\ref{second}) in spatial and temporal components, using the constraint condition (\ref{constraint}), and substituting the decomposition (\ref{decomp1} - \ref{decomp2}) we find two decoupled Lagrangians given by:
	
	\bea {\cal L}_{\varphi_3, \theta, \Gamma,\ \varphi_2}&=&-\frac{25}{2}\,\,\varphi_3\nabla^2\varphi_3+20\,\,\varphi_3\nabla^6\theta-10\,\,\varphi_3 \nabla^2\dot{\Gamma}-15\,\,\varphi_3\nabla^4\varphi_2 -8\,\,\dot{\theta}\nabla^8\dot{\theta}\nn\\
	&&-10\,\,\theta\nabla^{10}\theta-4\,\,\theta\nabla^6\dot{\Gamma}-2\,\,\dot{\Gamma} \nabla^2\dot{\Gamma}- 18\,\,\Gamma \nabla^4\Gamma-6\,\,\varphi_2\nabla^4\dot{\Gamma}-\frac{9}{2}\,\,\varphi_2\nabla^6\varphi_2 \label{first}\eea
	
	\no and
	
	\bea {\cal L}_{\gamma\psi\varphi_1}&=& -2\dot{\gamma}\nabla^2\dot{\gamma}-8\gamma\nabla^4\gamma+12\dot{\gamma}\nabla^4\dot{\psi}+24\gamma\nabla^6\psi-12\dot{\gamma}\nabla^4{\varphi}_1\nn\\
	&& -18\dot{\psi}\nabla^6\dot{\psi} -8\psi\nabla^8\psi+36\dot{\psi}\nabla^6{\varphi}_1-18\varphi_1\nabla^6\varphi_1.\label{secondt}\eea
	
	\no As one can observe the resulting lagrangians has several couplings among the scalar fields and such couplings are involving time derivatives. 
	
	In order to diagonalize the first lagrangian (\ref{first}) and eliminate any dynamical content coming from time derivatives, we proceed with the following set of transformations: 
	\bea \varphi_2&=& \tilde{\varphi}_2-\frac{5\tilde{\varphi}_3}{3\nabla^2}+\frac{2\ddot{\tilde{\varphi}}_3}{3\nabla^4}+\frac{\dot{\tilde{\Gamma}}}{\nabla^2}+\frac{2\ddot{\tilde{\theta}}}{3}\\
	\varphi_3&=& \tilde{\varphi}_3-\frac{2\ddot{\tilde{\varphi}}_3}{3\nabla^2}-\dot{\tilde{\Gamma}}-\frac{2\nabla^2\ddot{\tilde{\theta}}}{3}\\
	\Gamma&=&\frac{2\dot{\tilde{\varphi}}_3}{3\nabla^2}+\tilde{\Gamma}+\frac{2\nabla^2\dot{\tilde{\theta}}}{3}\\
	\theta&=&\frac{\tilde{\varphi}_3}{\nabla^4}+\tilde{\theta};\eea
	
	\no which take the original set of variables $\Phi_I \equiv (\varphi_2,\varphi_3,\Gamma,\theta)$ to the new one $\tilde{\Phi}_J\equiv(\tilde{\varphi}_2,\tilde{\varphi}_3,\tilde{\Gamma},\tilde{\theta})$ through the matrix ${\mathbb M}_{IJ}$ whose the determinant is unitary i.e: $\det {\mathbb M} =1$. In matrix form we have $\Phi_I={\mathbb M}_{IJ}\Phi_J$. After that, we obtain:
	
	\be \tilde{{\cal L}}_{\tilde{\varphi}_3,\tilde{\varphi}_2, \tilde{\theta},\tilde{\Gamma}}=10\,\,\tilde{\varphi}_3\nabla^2\tilde{\varphi}_3-10\,\,\tilde{\theta}\nabla^{10}\tilde{\theta}-18\,\,\tilde{\Gamma}\nabla^4\tilde{\Gamma}-\frac{9}{2}\,\,\tilde{\varphi_2}\nabla^6\tilde{\varphi}_2\label{dec1}\ee
	
	\no which clearly does not propagate any degrees of freedom. It remains now, to demonstrate that the lagrangian (\ref{secondt}) is also free of particle content, so, we have to diagonalize it also. 
	
	In (\ref{secondt}) we make the transformations:
	
	\bea \varphi_1 &=& \tilde{\varphi}_1-\frac{\dot{\tilde{\gamma}}}
	{3\nabla^2}+\frac{\dot{\tilde{\psi}}}{2},\\ 
	\gamma &=&  \tilde{\gamma}+\frac{3\nabla^2}{2}\tilde{\psi}\\
	\psi&=& \tilde{\psi}\eea
	
	\no taking the original variables $\Phi_I \equiv (\varphi_1,\gamma,\psi)$ to the new ones $\tilde{\Phi}_J\equiv(\tilde{\varphi}_1,\tilde{\gamma},\tilde{\psi})$ through the matrix ${\mathbb J}_{IJ}$ whose the determinant is also unitary. After that, substituting the new variables we obtain:
	\be \tilde{{\cal L}}_{\tilde{\varphi}_1, \tilde{\gamma},\tilde{\psi}}=-8\,\,\tilde{\gamma}\nabla^4\tilde{\gamma}+10\,\,\tilde{\psi}\nabla^8\tilde{\psi}-18\,\,\tilde{\varphi}_1\nabla^6\tilde{\varphi}_1.\label{dec2} \ee
	
	Therefore, once the lagrangians (\ref{dec1}) and (\ref{dec2}) are completely decoupled and evidently are not propagating any of their fields it can be guaranteed that the second order lagrangian given by (\ref{secondorderterm}) can in fact be used as a mixing term.
	
	\subsubsection{The third order mixing term}
	Despite the third order mixing term
	\be {\cal L}^{(3)}= \mathbb{C}_{\mu\nu\gamma\lambda}(\phi)\mathbb{G}^{\mu\nu\gamma\lambda}(\phi),\ee
	to be invariant under a bigger set of gauge symmetries, which would allow us to fix one more constraint equation, see the discussion in \cite{nges4}, we still use only the gauge condition (\ref{constraint}) for sake of simplicity. Notice that, this is allowed because we are fixing less constraint equations than the number of equations allowed by the gauge invariance. The opposite situation would be clearly forbidden, i.e. to fix more constraint equantions than the number allowed by the gauge invariance. 
	
	After substituting the decompositions (\ref{decomp1} - \ref{decomp2}) and the constraint equation, we obtain:
	\bea {\cal L}&=& 68\,\,\dot{\varphi}_3\nabla^2\dot{\gamma}+20\,\,\varphi_3\nabla^4\gamma-12\,\,\dot{\varphi}_3\nabla^4\dot{\psi}-60\,\,\varphi_3\nabla^6\psi+108\,\,\varphi_3\nabla^4\dot{\varphi}_1\nn\\
	&-&52\,\,\dot{\theta}\nabla^6\dot{\gamma}-40\,\,{\theta}\nabla^8{\gamma}+28\,\,\dot{\theta}\nabla^8\dot{\psi}+40\,\,{\theta}\nabla^{10}{\psi}-36\,\,\dot{\theta}\nabla^8\dot{\varphi}_1\nn\\
	&-& 24\,\,\dot{\varphi}_2\nabla^4\dot{\gamma}+12\,\,{\varphi}_2\nabla^6{\gamma}+24\,\,\dot{\varphi}_2\nabla^6\dot{\psi}+12\,\,{\varphi}_2\nabla^8{\psi}\nn\\
	&+&12\,\,\Gamma\nabla^4\dot{\gamma}-36\,\,\Gamma\nabla^6\dot{\psi}+48\,\,\dot{\Gamma}\nabla^4\dot{\varphi}_1+36\,\,\Gamma\nabla^6\varphi_1\eea
	which is a unique and completely coupled lagrangian. In order to decouple all the fields and eliminate the time derivatives, we will proceed with some rounds of rather technical transformations which obviously could be presented at once, but in order to be more didactical we have divided in sub-steps. The first round consists of:
	\bea \varphi_3&=&\bar{\varphi}_3-\frac{6\,\ddot{\bar{\varphi}}_3}{\nabla^2}+\frac{\nabla^4\bar{\theta}}{3}-2\,\dot{\bar{\Gamma}}\nn\\
	\varphi_2&=&\bar{\varphi}_2 +\frac{40\Phi_1}{3\nabla^2}\nn\\
	\theta&=&\bar{\theta}-\frac{14\Phi_1}{\nabla^4}\nn\\
	\Gamma&=& \bar{\Gamma}+\frac{3\dot{\bar{\varphi}}_3}{\nabla^2},
	\eea
	while $(\gamma,\psi,\varphi_1)\to (\bar\gamma,\bar\psi,\bar\varphi_1)$. Here, we have defined the combination $\Phi_1\equiv \ddot{\bar\varphi}_3/\nabla^2+\dot{\bar\Gamma}/3$.  This give us the following lagrangian free of the time derivative couplings   involving the pairs $(\varphi_3 , \varphi_1)$, $(\Gamma,\varphi_1)$ and $(\theta,\varphi_1)$:
	\bea {\cal \bar L}&=& -496\,\,\dot{\bar\varphi}_3\nabla^2\dot{\bar\gamma}+20\,\,\bar\varphi_3\nabla^4\bar\gamma-80\,\,\dot{\bar\varphi}_3\nabla^4\dot{\bar\psi}-60\,\,\bar\varphi_3\nabla^6\bar\psi\nn\\
	&-&\frac{88}{3}\,\,\dot{\bar\theta}\nabla^6\dot{\bar\gamma}-\frac{100}{3}\,\,{\bar\theta}\nabla^8{\bar\gamma}+24\,\,\dot{\bar\theta}\nabla^8\dot{\bar\psi}+20\,\,{\bar\theta}\nabla^{10}{\bar\psi}\nn\\
	&-& 24\,\,\dot{\bar\varphi}_2\nabla^4\dot{\bar\gamma}+12\,\,{\bar\varphi}_2\nabla^6{\bar\gamma}+24\,\,\dot{\bar\varphi}_2\nabla^6\dot{\bar\psi}+12\,\,{\bar\varphi}_2\nabla^8{\bar\psi}\nn\\
	&+&188\,\,\dot{\bar\Gamma}\nabla^4\bar{\gamma}+\frac{68}{3}\,\,\dot{\bar\Gamma}\nabla^6\bar{\psi}+36\,\,\bar\Gamma\nabla^6\bar\varphi_1.\eea
	
	\no Then, we proceed with a second round of transformations:
	\bea \bar\varphi_3&=&\bar{\bar\varphi}_3-\frac{\nabla^4\bar{\bar\theta}}{108}\nn\\
	\bar\varphi_2&=&\bar{\bar\varphi}_2-\frac{167\nabla^2{\bar{\bar\theta}}}{162}\nn\\
	\bar\varphi_1&=&\bar{\bar\varphi}_1 +\frac{47\ddot{\bar{\bar\gamma}}}{9\nabla^2}+\frac{17\dot{\bar{\bar\psi}}}{27}
	\eea
	\no while $(\bar\gamma,\bar\psi,\bar\theta,\bar{\bar\Gamma})\to(\bar{\bar\gamma},\bar{\bar\psi},\bar{\bar\theta},\bar{\bar\Gamma})$. After this transformation, one can show that:
	\bea {\cal \bar{\bar L}}&=& -496\,\,\dot{\bar{\bar\varphi}}_3\nabla^2\dot{\bar{\bar\gamma}}+20\,\,\bar{\bar\varphi}_3\nabla^4\bar{\bar\gamma}-80\,\,\dot{\bar{\bar\varphi}}_3\nabla^4\dot{\bar{\bar\psi}}-60\,\,\bar{\bar\varphi}_3\nabla^6\bar{\bar\psi}\nn\\
	&+&24\,\,\dot{\bar{\bar\varphi}}_2\nabla^4\dot{\bar{\bar\gamma}}+12\,\,{\bar{\bar\varphi}}_2\nabla^6{\bar{\bar\gamma}}+24\,\,\dot{\bar{\bar\varphi}}_2\nabla^6\dot{\bar{\bar\psi}}+12\,\,{\bar{\bar\varphi}}_2\nabla^8{\bar{\bar\psi}}\nn\\
	&+&\frac{413}{9}\,\,{\bar{\bar\theta}}\nabla^8{\bar{\bar\gamma}}+\frac{221}{27}\,\,{\bar{\bar\theta}}\nabla^{10}{\bar{\bar\psi}}+36\,\,\bar{\bar\Gamma}\nabla^6\bar{\bar\varphi}_1.\eea
	Which is now, free of time derivatives involving the pairs $(\theta,\gamma)$, $(\theta,\psi)$ and $(\Gamma,\gamma)$. Then, introducing the following set of transformations:
	\bea \bar{\bar\theta}&=&\tilde{\theta}+ \frac{36\Phi_2}{a\nabla^6}+\frac{108\Phi_3}{a\nabla^4}\nn\\
	\bar{\bar\gamma}&=&\tilde{\gamma}+\frac{221\nabla^2\tilde{\psi}}{3a}
	\eea
	while $(\bar{\bar\gamma},\bar{\bar\psi},\bar{\bar\varphi}_1,\bar{\bar\Gamma},\bar{\bar\theta})\to (\tilde{\gamma},\tilde{\psi},\tilde{\varphi}_1,\tilde{\Gamma},\tilde{\theta})$, where we have defined the numerical coefficient $a=413$ and the combinations $\Phi_2\equiv 5\nabla^2\tilde{\varphi}_3+124\ddot{\tilde{\varphi}}_3$ and $\Phi_3\equiv\nabla^2\tilde{\varphi}_2+2\ddot{\tilde{\varphi}}_2$. Such transformations take us to the lagrangian:
	\bea \tilde{ \cal{L}}&=&-\frac{1}{3a}(208736\,\,\dot{\tilde{\varphi}}_3\nabla^4\dot{\tilde{\psi}}+69920\,\,\tilde{\varphi}_3\nabla^6\tilde{\psi})\nn\\
	&+&\frac{1}{a}(8144\,\,\dot{\tilde\varphi}_2\nabla^6\dot{\tilde{\psi}}+5840\,\,\tilde{\varphi}_2\nabla^8\tilde{\psi})\nn\\
	&-&\frac{a}{9}\,\tilde{\theta}\nabla^8\tilde{\gamma}+36\,\tilde{\Gamma}\nabla^6\tilde{\varphi}_1\eea
	In order to eliminate by complete all time derivatives we implement:
	\bea \tilde{\varphi}_3&=&\hat{\varphi}_3+\frac{b}{a}\Phi_4\nn\\
	\tilde{\varphi}_2&=&\hat{\varphi}_2+\frac{13046}{1527}\hat{\varphi}_3-\frac{208736\,\,b}{1527\,\,a\nabla^2}\Phi_5\eea
	while $(\tilde{\gamma},\tilde{\psi},\tilde{\varphi}_1,\tilde{\Gamma},\tilde{\theta})\to(\hat{\gamma},\hat{\psi},\hat{\varphi}_1,\hat{\Gamma},\hat{\theta})$. Here, we have defined the numerical complicated coefficient $b=210217/13533120$ and the new combinations $\Phi_4\equiv 8144\,\ddot{\hat{\varphi}}_2-5840\,\nabla^2\hat{\varphi}_2$ and $\Phi_5\equiv 365\,\nabla^2\hat{\theta}-504\,\ddot{\hat{\theta}}$. This will take us to the following result:
	\be \hat{\cal L}=\frac{1}{b}\hat{\varphi}_3\nabla^6\hat{\psi}-\frac{a}{9}\hat{\theta}\nabla^8\hat{\gamma}+36\hat{\Gamma}\nabla^6\hat{\varphi}_1\ee
	\no where one can notice that we do not have any time derivatives in game but still some couplings between the fields. Such couplings can be easily eliminated by simple rotations defined by:
	\bea \Psi_1&=&\frac{\sqrt{2}}{2}(\hat{\Gamma}-\hat{\varphi})\quad;\quad \Psi_2=\frac{\sqrt{2}}{2}(\hat{\Gamma}+\hat{\varphi})\nn\\
	\Psi_3&=&\frac{\sqrt{2}}{2}(\hat{\theta}-\hat{\gamma})\quad;\quad \Psi_4=\frac{\sqrt{2}}{2}(\hat{\theta}+\hat{\gamma})\nn\\
	\Psi_5&=&\frac{\sqrt{2}}{2}(\hat{\varphi}_3-\hat{\psi})\quad;\quad \Psi_6=\frac{\sqrt{2}}{2}(\hat{\varphi}_3+\hat{\psi})\eea
	which finally give us the completely decoupled lagrangian:
	
	\be {\cal{L}}= \frac{1}{2b}(\Psi_6\nabla^6 \Psi_6-\Psi_5\nabla^6\Psi_5)-\frac{a}{18}(\Psi_4\nabla^8 \Psi_4-\Psi_3\nabla^8\Psi_3)+18(\Psi_2\nabla^6 \Psi_2-\Psi_1\nabla^6\Psi_1),\ee
	
	\no which is evidently free of particle content. It is remarkable that all the transformations we have done have unitary determinants as we have observed in the case of the second-order term, guaranteeing that the descriptions in terms of the new variables are canonically equivalent to the original ones. Besides, it is obvious that the complicated numerical coefficients $a$ and $b$ can be absorbed by simple redefinitions of the fields which makes the result more elegant. Besides, notice that an alternative way of analysis, based on the identification of gauge invariant objects could also be used to deal with this demonstration, such an approach has successfully been used in the spin-1, 2, 3, and 4 cases in \cite{denis} for example.
	
	Now that we have a safe first order self-dual model, free of ghosts, describing correctly a massive spin-4 mode in $D=2+1$ and three trivial mixing terms we are in position to construct the master action, this will be done in the next subsection:
	\subsection{The master action}
	 Using the first  order self-dual model (\ref{SD(1)}) as our starting point we proceed  by plugging to it,  three mixing terms, and this is what we call the master action: 
	
	\bea S_M&=& \int \left[\frac{m}{2}\omega\cdot d\omega +\frac{m^2}{2}(\omega^2)+c_1\,(\omega-g)\cdot
	d(\omega-g) +c_2 \,(h-g)\cdot d \Omega(h-g)\right.\nn\\
	&+& \left. c_3\, \Omega(s-h)\cdot d \Omega (s-h)\right] + m^2\int d^3x\,\,\omega_{\mu\nu}U^{\mu\nu}+
	S^1_{aux}[U,H,V].\label{mestra}\eea

	\no Where the subscript $M$ refers to the Master action while {\it aux} stands for auxiliary. In the action (\ref{mestra}), the terms preceded by the coefficients $c_1$, $c_2$, and $c_3$ are the so-called mixing terms, while such coefficients are { \it a priori} arbitrary. Of course, if $c_1=c_2=c_3 =0$ we have exactly the first order self-dual model by \cite{aragone2}. While $\omega$ is the spin-4 field, $g$, $h$, and $s$ are introduced to be used in the interpolation process and have the same symmetry in their indices as $\omega$.  During the interpolation process, the auxiliary action will also change, that is why we use the superscript $1$ denoting the first auxiliary action in the process. 
	
	To verify the quantum equivalence of the self-dual descriptions being interpolated by the master action (\ref{mestra}), we add a source term $j_{\mu(\beta\gamma\lambda)}$ to the spin-4 field and define the generating functional.
	
	\be W_M[j]= \int\,\, {\cal D}\omega\, {\cal D}g\,{\cal D}h\,{\cal D}U\,{\cal D}H\,{\cal D}V\, \,\,exp\,\,i\left (S_M+\int d^3x \,j_{\mu(\beta\gamma\lambda)}\omega^{\mu(\beta\gamma\lambda)}\right).\label{GF}\ee
	
	\no The first equivalence one can check is done by means of the following shifts on the master action (\ref{mestra}), in this order: $s\to s +h$,  $h\to h + g$ and $g\to g+ \omega$. With such shifts the three mixing terms get completely decoupled and once they are free of particle content, they can be functionally integrated out. Then, one conclude by the equivalence of the correlation functions of the master action (\ref{mestra}) with the first order self-dual model here represented by the subscript $SD(1)$.
	
	\be \langle \omega_{\mu_1(\alpha_1\beta_1\gamma_1)}(x_1)...\omega_{\mu_N(\alpha_N\beta_N\gamma_N)}(x_N)\rangle_{M}=\langle \omega_{\mu_1(\alpha_1\beta_1\gamma_1)}(x_1)...\omega_{\mu_N(\alpha_N\beta_N\gamma_N)}(x_N)\rangle_{SD(1)}.\ee
	\no we have then verified that the suggested master action has the quantum spectrum of the self-dual model and, so is free of ghosts, is free of the propagation of lower spins and propagate a unique spin-4 mode in $D=2+1$ dimensions. In the next section we demonstrate that, starting with the very same master action one can obtain higher derivative descriptions which are quantum equivalents to the good (free of ghosts) first order self-dual model.
	\section{Higher derivative quantum equivalent descriptions} 
	\subsection{Interpolating with $SD(2)$}
	On the other hand, back in the master action (\ref{mestra}), if we perform just the first two shifts $s\to s+h$ and  $h\to h+g$ we can decouple only the mixing terms preceded by $c_2$ and $c_3$, while the remaining one, must be opened in order to take us to the following intermediary step:
	
	\be S_M[j]= \int \, \left\lbrack \frac{m^2}{2}{(\omega^2)}-\frac{m}{2}g\cdot dg\right\rbrack + \int d^3x\,\, \omega_{\mu(\beta\gamma\lambda)}\tilde{g}^{\mu(\beta\gamma\lambda)}+S^1_{aux},\label{smg}\ee
	
	\no where we have chosen $c_1=-m/2$. Besides, we have defined:
	\be \tilde{g}^{\mu(\beta\gamma\lambda)}\equiv -m\xi^{\mu(\beta\gamma\lambda)}+\frac{m^2}{3}f^{\mu(\beta\gamma\lambda)}(U)+j^{\mu(\beta\gamma\lambda)},\label{gtil}\ee
	\no with the partially symmetric-traceless combination of the auxiliary fields $\tilde{U}^{(\beta\gamma)}$ explicitly given by:
	\be f^{\mu(\beta\gamma\lambda)}(\tilde{U}) \equiv \eta^{\mu\beta}\tilde{U}^{(\gamma\lambda)}+\eta^{\mu\gamma}\tilde{U}^{(\beta\lambda)}+\eta^{\mu\lambda}\tilde{U}^{(\gamma\beta)}- \frac{2}{5} (\eta^{\beta\gamma}\tilde{U}^{(\mu\lambda)}+\eta^{\beta\lambda}\tilde{U}^{(\mu\gamma)}+\eta^{\gamma\lambda}\tilde{U}^{(\mu\beta)}).\ee
	
	\no Since in (\ref{smg}), we have a quadratic and a linear term on $\omega$ one can functionally integrate over it in (\ref{GF}), obtaining:
	\be S_{M}[j]= -\frac{m}{2}\int\, g\cdot dg - \frac{1}{m^2}\int \,d^3x\, \tilde{g}_{\mu(\beta\gamma\lambda)}\Omega^{\mu(\beta\gamma\lambda)}(\tilde{g})+S^1_{aux}.\label{gom}\ee
	
	\no After substituting back $\tilde{g}$ from (\ref{gtil}) in (\ref{gom}) we finally have:
	\bea S_{SD(2)}[j]= \int \,\left\lbrack -\frac{m}{2}g\cdot dg+ g \, \cdot d\Omega(g)-\frac{m}{4}g\cdot d f +j_{\mu(\beta\gamma\lambda)}F^{\mu(\beta\gamma\lambda)}(g,U)+{\cal O}(j^2)\right\rbrack +S^2_{aux};\label{sd2}\nn\\\eea 
	
	\no where $f$ is an abbreviation to $f(U)$. The model (\ref{sd2}), is exactly the second order self-dual model $SD(2)$ with all the new corrections in the auxiliary fields and in the linking terms. Explicitly, the term $-m\,g\cdot d f/4 = 3m \,\xi_{\mu\beta}\tilde{U}^{\mu\beta}/4$ and the corrected auxiliary action, $S^2_{aux}$, is given by:
	
	\be S^2_{aux}= S^1_{aux}-\frac{21m^2}{40}\int\, d^3x\, \tilde{U}_{(\alpha\beta)}\tilde{U}^{(\alpha\beta)}.\ee
	\no Such results can be compared with those we have obtained in \cite{nges4} through the Noether Gauge Embedment technique \footnote{Observe we have a little mistake in \cite{nges4} where the coefficient $21/40$ has been written as $11/40$.}. 
	Notice also, that we have gained a new kind of coupling with the source term, which is now given in terms of the dual (gauge invariant) combination $F^{\mu(\beta\gamma\lambda)}$ which establish a dual map between the field of $SD(1)$ and the gauge invariant combination $F$ of $SD(2)$, this is given by:
	
	\be \omega^{\mu(\beta\gamma\lambda)} \longleftrightarrow F^{\mu(\beta\gamma\lambda)}(g,U)=\frac{2}{m}\Omega^{\mu(\beta\gamma\lambda)}(g)-\frac{1}{4}f^{\mu(\beta\gamma\lambda)}(U),\ee
	
	\no On the other hand, by using the dual map, one can recover the equations of motion of either the first-order self-dual model ($SD(1)$) or the second-order self-dual model ($SD(2)$) from one another, verifying the equivalence of the models at the classical level. To demonstrate the equivalence at the quantum level, we take the derivative of the generating functional with respect to the source in both the master action (\ref{GF}) and the second-order self-dual model (\ref{sd2}), which results in equivalent correlation functions.
	
	\be \langle \omega_{\mu_1(\alpha_1\beta_1\gamma_1)}(x_1)...\omega_{\mu_N(\alpha_N\beta_N\gamma_N)}(x_N)\rangle_{M}=\langle F_{\mu_1(\alpha_1\beta_1\gamma_1)}(x_1)...F_{\mu_N(\alpha_N\beta_N\gamma_N)}(x_N)\rangle_{SD(2)}+ C.T\ee
	
	\no where $C.T$ stands for Contact Terms, which comes from the quadratic terms on the source. They are proportional to the Kronecker and Dirac deltas and do not affect the quantum equivalence at all.
	
	\subsection{Interpolating with SD(3)}

   As we have seen, the master action is equivalent to the second order self-dual model $SD(2)$ up to contact terms, then, considering (\ref{sd2}) as our new starting point and adding back the mixing term proportional to $c_2$. we have:
   
   \bea S_M[j]&=& \int \,\left\lbrack -\frac{m}{2}g\cdot dg+ g \, \cdot d\Omega(g)-\frac{m}{4}g\cdot d f+ c_2\,(h-g)\cdot d \Omega(h-g)\right.\nn\\
   &+& \left.j_{\mu(\beta\gamma\lambda)}F^{\mu(\beta\gamma\lambda)}(g,U)+{\cal O}(j^2)\right\rbrack +S^2_{aux}\label{sd222}\nn\\\eea 
  
 \no  choosing $c_2=-1$, the action (\ref{sd222}) can be rearranged, after opening the mixing term, as:
  \bea S_M[j]&=& \int \,\left\lbrack -\frac{m}{2}g\cdot dg - g \, \cdot d C- h\cdot d \Omega(h)-\frac{1}{4}j_{\mu(\beta\gamma\lambda)}f^{\mu(\beta\gamma\lambda)}(U)+{\cal O}(j^2)\right\rbrack +S^2_{aux}\label{sd22}\nn\\\eea 
  
  \no where we have defined:
  
  \be C^{\mu(\beta\gamma\lambda)}\equiv\frac{2}{m^2}\Omega^{\mu(\beta\gamma\lambda)}(j)-\frac{2}{m}\Omega^{\mu(\beta\gamma\lambda)}(h)+\frac{1}{4}f^{\mu(\beta\gamma\lambda)}(U).\ee
  
  \no Once we have a quadratic and a linear term on $g$ in (\ref{sd22}) we can rewrite the master action as:
  
  \bea S_M[j]&=& \int \,\left\lbrack -\frac{m}{2}(g+C)\cdot d(g+C) +\frac{m}{2} C \, \cdot d C- h\cdot d \Omega(h)-\frac{1}{4}j_{\mu(\beta\gamma\lambda)}f^{\mu(\beta\gamma\lambda)}(U)+
  {\cal O}(j^2)\right\rbrack\nn\\&+&S^2_{aux}.\label{sd2222}\eea 
  
  \no  It is easy to see that the shift $g\to g+C$ completely decouple the trivial Chern-Simons like term, which then can be functionally integrated out. Then, after substituting back the combination $C$ in (\ref{sd2222}) we have:
 \bea S_{SD(3)}[j]&=& \int \,\left\lbrack - h\cdot d \Omega(h)+\frac{2}{m}\Omega(h)\cdot d \Omega(h)-\frac{1}{2}f\cdot d\Omega(h)+\tilde{\Omega}_{\mu(\beta\gamma\lambda)}(j)H^{\mu(\beta\gamma\lambda)}(h,U)+
 {\cal O}(j^2)\right\rbrack\nn\\
 &+& S_{aux}^3,\label{sd3}\eea
 
 \no which is precisely the third order self-dual model $SD(3)$ we have obtained before in \cite{nges4}. The new auxiliary action is given by:
 \be S_{aux}^{3}=S_{aux}^2+ \frac{21m}{80} \int d^3x\, \, \tilde{U}_{\mu\beta}E^{\mu}_{\,\,\,\,\gamma}\tilde{U}^{\gamma\beta}.\ee
 Here, the source is coupled with the dual $H^{\mu(\beta\gamma)}$ given by:
 \be H^{\mu(\beta\gamma)}(h,U)= \frac{1}{2m}E^{\mu}_{\,\,\alpha}f^{\alpha(\beta\gamma\lambda)}(U)-\frac{4}{m^2}E^{\mu}_{\,\,\alpha}\Omega^{\alpha(\beta\gamma\lambda)}(h)-\frac{2}{3}f^{\mu(\beta\gamma\lambda)}(U).\ee
 through the combination of the source $\tilde{\Omega}(j)$, and then the dual map with the previous self-dual models is given by:
 
 \be \omega^{\mu(\beta\gamma\lambda)} \longleftrightarrow \tilde{\Omega}^{\mu(\beta\gamma\lambda)}[H(h,U)].\ee
 
 Which, after deriving with respect to the source term from (\ref{mestra}) and (\ref{sd3}) give us the correspondence of the correlation functions:

 \be \langle \omega_{\mu_1(\alpha_1\beta_1\gamma_1)}(x_1)...\omega_{\mu_N(\alpha_N\beta_N\gamma_N)}(x_N)\rangle_{M}=\langle \tilde{\Omega}(H)_{\mu_1(\alpha_1\beta_1\gamma_1)}(x_1)...\tilde{\Omega}(H)_{\mu_N(\alpha_N\beta_N\gamma_N)}(x_N)\rangle_{SD(3)}+ C.T.\ee
 
 \no In the next section, we will simplify our notation by shifting from partially symmetric to totally symmetric notation, making it easier to interpolate with the fourth-order self-dual model. This change is not necessary to observe the equivalence, but it helps to avoid an excess of symbols and to make the calculations more straightforward. Although the source term is coupled to the auxiliary fields and the spin-4 fields in a complex manner, we will simplify it by redefining the linear, partially symmetric combination $\tilde{\Omega}(j)$ in (\ref{sd3}) to simply $J$. For simplicity, we will also neglect the coupling of the source to the auxiliary sector and focus only on its coupling to the totally symmetric fields.

	\subsection{ $SD(3)$ to  $SD(4)$ - totally symmetric approach}
	Back to the third order self-dual model, taking advantage that we have verified in the last section its equivalence with the master action, we finally add the last mixing term preceded by the coefficient $c_3$, which then give us the following master action:
	\bea S_{M}[j]&=& \int \,\left\lbrack - h\cdot d \Omega(h)+\frac{2}{m}\Omega(h)\cdot d \Omega(h)-\frac{1}{2}f\cdot d\Omega(h)+  c_3\, \Omega(s-h)\cdot d \Omega (s-h)\right.\nn\\
	&+& \left. \frac{4}{m^2}J_{\mu(\beta\gamma\lambda)}H^{\mu(\beta\gamma\lambda)}(h)+
	{\cal O}(J^2)\right\rbrack+ S_{aux}^3.\label{sd32}\eea
	
	Applying the decomposition introduced in (\ref{deco}) to the fields $h$, $s$ and to the source $J$ and choosing $c_3=-2/m$ we have:
	\bea S_{M}&=& \int d^3x \left[\frac{1}{2} \phi_{\mu\nu\lambda\beta}\,\mathbb{G}^{\mu\nu\lambda\beta}(\phi) +\frac{1}{8m}\mathbb{C}_{\mu\nu\gamma\lambda}(\phi)\mathbb{G}^{\mu\nu\gamma\lambda}(\phi) -\frac{1}{8m}\mathbb{C}_{\mu\nu\gamma\lambda}(\sigma-\phi)\mathbb{G}^{\mu\nu\gamma\lambda}(\sigma-\phi)\right.\nn\\
	&+&\left.{\cal U}_{\mu\nu\lambda\beta}\mathbb{G}^{\mu\nu\lambda\beta}(\phi)-\frac{2}{m^2} {\cal J}_{\mu\nu\lambda\beta}\mathbb{G}^{\mu\nu\lambda\beta}(\phi)\right]+ S_{aux}^3,\label{sds}\eea
	
    \no where we have defined ${\cal U}_{\mu\nu\lambda\beta}\equiv 3\eta_{(\mu\nu}\tilde{U}_{(\lambda\beta))} /40$. The fields $\phi$ and $\sigma$ corresponds respectively to the totally symmetric double traceless parts of $h$ and $s$. The source ${\cal J}$ is the totally symmetric part of the redefined source $J$ and it is perfectly coupled to the field $\phi$ through the Einstein tensor. 
    
    Making the shift $\sigma \to \sigma +\phi$, it is obvious that the master action becomes the third order self-dual model in its totally symmetric version. On the other hand, by opening the mixing term in (\ref{sds}), we have:

\bea S_{M}&=& \int d^3x \left[-\frac{1}{2} \psi_{\mu\nu\lambda\beta}\,\mathbb{G}^{\mu\nu\lambda\beta}(\psi) -\frac{1}{8m}\mathbb{C}_{\mu\nu\gamma\lambda}(\sigma)\mathbb{G}^{\mu\nu\gamma\lambda}(\sigma) \right]+ S_{aux}^3,\label{sdsin}\eea

\no where we have defined $\psi$, given by:

\be \psi_{\mu\nu\lambda\beta}\equiv \frac{1}{4m}\mathbb{C}_{\mu\nu\lambda\beta}(\sigma)+ {\cal U}_{\mu\nu\lambda\beta}-\frac{2}{m^2}{\cal J}_{\mu\nu\lambda\beta} ,\label{psi}\ee

\no and made the shift $\phi \to \phi -\psi$. By substituting back the field $\psi$ from (\ref{psi}) in (\ref{sdsin})  we finally have:
\bea S_{SD(4)}&=& \int d^3x \,\,\left[-\frac{1}{8m}\mathbb{C}_{\mu\nu\gamma\lambda}(\phi)\mathbb{G}^{\mu\nu\gamma\lambda}(\phi)-\frac{1}{32m^2}\mathbb{C}_{\mu\nu\gamma\lambda}(\phi)\mathbb{G}^{\mu\nu\gamma\lambda}(\mathbb{C})-\frac{1}{4m}\mathbb{C}_{\mu\nu\gamma\lambda}(\phi)\mathbb{G}^{\mu\nu\gamma\lambda}({\cal U})\right.\nn\\
&-& \left.\frac{2}{m^2}{\cal J} _{\mu\nu\lambda\beta}\mathbb {G}^{\mu\nu\lambda\beta}\left(\chi\right)+{\cal O}({\cal J}^2)\right]+ S_{aux}^4 ; \label{sd4s}\eea

	\no where we have defined $\chi\equiv -{{\mathbb C}(\sigma)}/{4m}-{\cal U}$.  The corrected auxiliary action is given by:
	\be S_{aux}^4 = S_{aux}^3 - \int d^3x\,\, \frac{1}{2}{\cal {U}}_{\mu\nu\gamma\lambda}\mathbb{G}^{\mu\nu\gamma\lambda}({\cal{U}}).\ee
	
	Taking functional derivatives with respect the source from (\ref{sds}) and (\ref{sd4s}) we have finally the equivalence of the correlation functions:
	\be \langle \mathbb{G}_{\mu_1\alpha_1\beta_1\gamma_1}(\phi)...\mathbb{G}_{\mu_N\alpha_N\beta_N\gamma_N}(\phi)\rangle_{M}=\langle \mathbb{G}_{\mu_1\alpha_1\beta_1\gamma_1}(\chi)...\mathbb{G}_{\mu_N\alpha_N\beta_N\gamma_N}(\chi)\rangle_{SD(4)}+ C.T.\ee
	
	In the notation we have used here, based on \cite{deserdamour}, the Einstein tensor is second order in derivatives, however in \cite{Henneaux} the authors have done a study on the conformal geometry of higher spin bosonic gauge fields in three spacetime dimensions where the Einstein tensor is proportional to the Riemann tensor, and in this case for a rank-$s$ field the Einstein tensor is of order $s$ in derivatives.

	\section{Conclusion}
	In this paper, we have been using the master action technique to show that in $D=2+1$ dimensions, the massive spin-4 particle can be described by four quantum equivalent self-dual descriptions, abbreviated as $SD(i)$ with $i=1,2,3,$ and $4$ indicating the order in derivatives of each model. The equivalence is confirmed by comparing the correlation functions, up to contact terms. The first two self-dual descriptions, $SD(1)$ and $SD(2)$, were introduced by Aragone and Khoudeir \cite{aragone2}, and are written in terms of partially symmetric fields $\omega_{\mu(\beta\gamma\lambda)}$ using the frame-like approach from Vasiliev \cite{Vasiliev}. On the other hand, $SD(3)$ and $SD(4)$ can be written in terms of totally symmetric double traceless fields, enabling us to introduce a geometrical description based on \cite{deserdamour}.
	
	At the moment, we are working on understanding the relationship between the notation used by Aragone and Khoudeir and the notation used by $[18]$. This appears to be well-established in the work of those authors; however, even they acknowledge that the notation, while possible to generalize to arbitrary spin (fermions and bosons), is considerably more complicated than the original notation. In this regard, it is worth reading the text below equation (49) in their work. It is worth mentioning that in our case, the situation seems to be even more difficult when dealing with higher-order derivative models where topologically structured terms are common, i.e., akin to Chern-Simons terms. Studying this relationship is interesting in terms of understanding how to generalize our results to the arbitrary spin-s case, where the main field would be a 1-form represented by $\omega_{\mu(\alpha_1...\alpha_s)}$. Alternatively, the transition to the differential forms formalism (indeed) may also be quite promising. This includes the extension of our results to curved space. This would be done by the following identification: First, we would distinguish between world indices $\mu$ and Lorentz indices $b_1 ...b_s$, and thus we would have $\omega_{\mu(b_1...b_s)}$, where the indices would then be interconnected by the dreibein $e^{\mu}_{a}$ \footnote{In the flat case, the dreibein reduces to the Kronecker delta} i.e., $\omega_{a(b_1...b_s)} = e^{\mu}_a\,\,\omega_{\mu(b_1...b_s)}$. At the same time, the contractions that we use to form our terms can in some cases be interpreted using the notation of differential forms and exterior derivatives, such as the first-order derivative term:
	\be
	\underbrace{\int \,\, \omega\cdot d\omega}_{our \,\, notation} \equiv \underbrace{\int d^3x\,\,\epsilon^{\mu\nu\alpha}\omega_{\mu(\beta\gamma\lambda)}\p_{\nu}\omega_{\alpha}^{\s(\beta\gamma\lambda)}}_{explicit}\equiv \underbrace{\int \,\, \omega\wedge d\omega}_{exterior}.
	\ee
	Unfortunately, with respect to the other terms we are working with, especially those involving the symbol $\Omega$, it is not clear to us how to encode them in terms of the exterior notation. We know from the definition of $\Omega$ given by (\ref{OM1}) that $\Omega \cong \,\wedge\, d \omega + ...$, but note that it is a symmetric and traceless combination of $\wedge\, d\, \omega$, and it is not trivial to see how we can deal with that in this formalism.
	
	It is worth mentioning, however, that if we were working with a field of arbitrary rank $\omega_{\mu(\alpha_1...\alpha_s)}$, it appears that we would be able to define $\xi_{\mu(\alpha_1...\alpha_s)}$ and consequently $\Omega_{\mu(\alpha_1...\alpha_s)}$, which would then have the following appearance:
	\bea
	\Omega_{\mu(\alpha_1...\alpha_s)} &=& \xi_{\mu(\alpha_1...\alpha_s)}- \underbrace{c_1(\xi_{\alpha_1(\mu...\alpha_s)}+...+\xi_{\alpha_1(\mu...\alpha_s)})}_{s \,\, terms}\nn\\
	&+&\underbrace{c_2 (\eta_{\mu\alpha_1}\tilde{\xi}_{\alpha_2...\alpha_s}+...+\eta_{\mu\alpha_s}\tilde{\xi}_{\alpha_2...\alpha_{s-1}})}_{s\,\, terms}\nn\\
	&+&\underbrace{c_3 (\eta_{\alpha_1\alpha_2}\tilde{\xi}_{\mu...\alpha_s}+...+\eta_{\alpha_s\alpha_{s-1}}\tilde{\xi}_{\mu...\alpha_{s-1}})}_{s(s-1)/2\,\, terms}
	\eea
	with $c_1$, $c_2$, and $c_3$ coefficients to be determined to ensure the symmetry properties of the symbol $\Omega$ (symmetric and tracelless in the Lorentz indices) and at the same time in order to reproduce the Fronsdal kinetic term after contraction with $\xi_{\mu(\alpha_1...\alpha_s)}$, i.e., $\xi_{\mu(\alpha_1...\alpha_s)}\Omega^{\mu(\alpha_1...\alpha_s)}$. We inform that this corresponds to work in progress.
	
	In conclusion, this work has demonstrated that the massive spin-4 particle in $D=2+1$ dimensions can be described by four quantum equivalent self-dual descriptions, each with a different order in derivatives (first, second, third and fourth order in derivatives models). The equivalence we previously demonstrated in \cite{nges4} is verified only at the classical level, here we demonstrate that it persists in the quantum level.
	
	The reader may wonder why the master method has worked in various situations. The answer is simple: during the interpolation process between the self-dual models, after appropriate shifts, we can decouple the so-called mixing terms step by step. Since these terms do not have physical content (as demonstrated in Section 3.2), we can conclude that they can be dispensed with, establishing the equivalence between correlation functions, all supported by the correlation functions of the first-order self-dual model, which, as we checked in Section 3.1, accurately describes a propagating mode of spin-4. Therefore, we can argue that the sequence of models is indeed quantum equivalent.
	
	The process of constructing the master action required the identification of mixing terms. These terms must not contain any particle content, as they will be decoupled from the action during the interpolation process and can be functionally integrated out. In this work, three mixing terms were used: a first-order Chern-Simons like term (\ref{csl}), a second-order Einstein-Hilbert like term (\ref{ehl}), and a third-order topologically Chern-Simons term (\ref{tcs}). While the particle content of the first term is easy to verify, the same is not true for the other two terms. We analyzed their particle content using the helicity decomposition method, which was employed in previous works such as \cite{deserh, andringa, townsendh}. Our analysis involved decomposing the symmetric double tracelles fields and using a gauge fixing condition at the level of the action. In line with \cite{moto1} and \cite{moto2}, we showed that our gauge condition is complete, meaning the gauge parameter can be inverted unambiguously in terms of the field itself. We then carried out a series of technical transformations, resulting in the Lagrangians being diagonalized and free of time derivatives, which would have otherwise implied dynamical content. 
	
	Notice that, one could ask about the particle content of the fourth order in derivatives term emerged in the action of $SD(4)$ in (\ref{sd4s}), $\sim \mathbb{C}_{\mu\nu\gamma\lambda}(\phi)\mathbb{G}^{\mu\nu\gamma\lambda}(\mathbb{C})$, and if it could be used as a mixing term which would take us to a new self-dual description, but, at this point we argue that, since we do not have found any gauge symmetry in the work \cite{nges4} which could take us to the fifth order self-dual model, we strongly believe that such fourth order term has indeed particle content, i.e it is non trivial.
	
	In this paper, we have made a contribution to the ongoing discussion about higher derivative descriptions of spin-4 self-dual models. Our work is different from previous studies, such as \cite{denis} and \cite{kuzenko}. In \cite{denis}, the authors explored seventh and eighth order models in derivatives, which were described without the use of auxiliary fields and in terms of totally symmetric but not double traceless fields. On the other hand, \cite{kuzenko} focused on exploring the same models but in a maximally symmetric curved spacetime. Our work, while still addressing the topic of spin-4 self-dual models, offers a distinct perspective and contribution to the field.

	\section{Acknowledgements}
	
	The authors would like to thank Prof. Denis Dalmazi for useful discussion regarding the helicity decomposition method. The authors also would like thank Prof. Nicolas Boulanger for suggesting the interesting references \cite{XBek1, XBek2, Xbek3}.
	
	\section{Data Availability Statement}
	
	No associated data in the manuscript.

\end{document}